\documentclass[reprint,superscriptaddress,amsmath,amssymb,aps,pre,floatfix]{revtex4-1}

\usepackage{graphicx}% Include figure files
\usepackage{dcolumn}% Align table columns on decimal point
\usepackage{bm}% bold math
\usepackage{color}
\usepackage{bbold}
\usepackage{soul}

\usepackage{amsmath,amssymb}
\usepackage{dutchcal}
\newcommand{\cQ}{\mathcal{q}}
\newcommand{\cP}{\mathcal{p}}

\newcommand{\newc}{\newcommand}
\newc{\beq}{\begin{equation}}
\newc{\eeq}{\end{equation}}
\newc{\kt}{\rangle}
\newc{\br}{\langle}
\newc{\beqa}{\begin{eqnarray}}
\newc{\eeqa}{\end{eqnarray}}
\newc{\longra}{\longrightarrow}

\newc{\KRM}{{\textcolor{blue}{removed text by KR~  } }}
\newc{\KRerase}[1]{\textcolor{blue}{\st{#1}}}

\newcommand\KR[1]{{\textcolor{blue}{#1}}}
\newc{\PS}[1]{\textcolor{red}{#1}}
\newc{\PSerase}[1]{\textcolor{red}{\st{#1}}}

\begin{document}

\title{Post-Ehrenfest many-body quantum interferences in ultracold atoms far-out-of-equilibrium}

\author{Steven Tomsovic}
\affiliation{Institut f\"ur Theoretische Physik, Universit\"at Regensburg, 93040 Regensburg, Germany}
\affiliation{Department of Physics and Astronomy, Washington State University, Pullman, WA~99164-2814}

\author{Peter Schlagheck}
\affiliation{CESAM research unit, University of Liege, 4000 Li\`ege, Belgium}

\author{Denis Ullmo}
\affiliation{Laboratoire de Physique Th\'eorique et Mod\`eles Statistiques, Universit\'e Paris Sud 
        91405, Orsay, France}

\author{Juan-Diego Urbina}
\affiliation{Institut f\"ur Theoretische Physik, Universit\"at Regensburg, 93040 Regensburg, Germany}

\author{Klaus Richter}
\affiliation{Institut f\"ur Theoretische Physik, Universit\"at Regensburg, 93040 Regensburg, Germany}

\date{\today}

\begin{abstract}

Far out-of-equilibrium many-body quantum dynamics in isolated systems necessarily generate
interferences beyond an Ehrenfest time scale, where quantum and classical expectation values
diverge.  Of great recent interest is the role these interferences play in the spreading of quantum information across the many degrees of freedom, i.e.~scrambling.  Ultracold atomic gases provide a promising setting to explore these phenomena.  Theoretically speaking, the heavily-relied-upon truncated Wigner approximation leaves out these interferences.  We develop a semiclassical theory which bridges classical and quantum concepts in many-body bosonic systems and properly incorporates such missing quantum effects.  For mesoscopically populated Bose-Hubbard systems, it is shown that this theory captures post-Ehrenfest quantum interference phenomena very accurately, and contains relevant phase information to perform many-body spectroscopy with high precision.

\end{abstract}

\pacs{}

\maketitle

The tremendous progress that has been achieved experimentally with quantum gases in optical lattices is leading to a vast exploration of new many-body physics phenomena~\cite{Bloch08}.  The pioneering works in this context mainly focused on ground-state properties~\cite{Greiner02a}, whereas more recent experiments are exploring dynamical processes in far-from-equilibrium settings triggered by a sudden or continuous parameter variation in the trapping configuration.  Important examples for bosonic atoms include features such as tunneling~\cite{Albiez05, Lignier07, Folling07}, transport~\cite{Salger09}, Landau-Zener transitions~\cite{Chen11}, relaxation~\cite{Trotzky12}, thermalization~\cite{Kaufman16}, and many-body localization~\cite{Choi16}. The experiments pose highly demanding challenges for numerical state-of-the-art simulations~\cite{Trotzky12, Choi16}, which underlines their possible role as quantum simulators.  

The above studies were most often  concerned with either a microscopically low or a macroscopically large number of atoms per site (i.e.~$O(1)$ or $>100$, respectively).  A new experimental regime is emerging with \emph{mesoscopic} populations of lattice sites with a few tens of atoms per site~\cite{Esteve08, Gross10}.  These mesoscopically populated lattices are expected to reveal interesting many-body physics due to the interplay between intrasite correlation and intersite tunneling effects~\cite{Chen11}.  More specifically, they allow one to probe the crossover from a classical mean-field regime, where the evolution of the Bose gas is well described by the Gross-Pitaevskii equation~\cite{Pitaevskii03} optionally in combination with a Bogoliubov ansatz~\cite{Castin98}, to a quantum correlated regime in which the mean-field approximation breaks down.

A relevant time scale that characterizes this crossover in a non-equilibrium context is the
Ehrenfest time, which is determined by the divergence of quantum and classical dynamics as time increases ~\cite{Ehrenfest27, Berman78}, and which marks the limit of validity of the mean-field ansatz.  Beyond this time scale, quantum interference effects become relevant, and give rise to significant physical phenomena ranging from the more spectacular, e.g. many-body localization~\cite{Gornyi05,Basko06} or quantum revivals~\cite{Greiner02b}, to the more subtle, e.g. coherent backscattering in Fock space~\cite{Engl14b}.  In a related perspective, the Ehrenfest time can be regarded as a delay time for the onset of quantum effects in quantum chaotic (many-body) systems~\cite{Larkin69}, and has recently attracted enormous attention.  The spreading of quantum information across degrees of freedom of a many-body system, commonly referred to as scrambling~\cite{Sekino08}, is expected to be governed by the Ehrenfest time.  Of particular note are out-of-time-ordered correlators~\cite{Shenker14}, representing the sensitivity of a time-evolving quantum observable to an initial perturbation.  They quantify scrambling and exhibit distinct deviations from the classical exponential growth behavior post-Ehrenfest~\cite{Rozenbaum17}.   Ultracold Bose gases within mesoscopically populated lattices therefore provide a promising setting to explore scrambling effects under well-controlled conditions~\cite{Yao16}.

\begin{figure*}[tbh]
\includegraphics[width=17.5 cm]{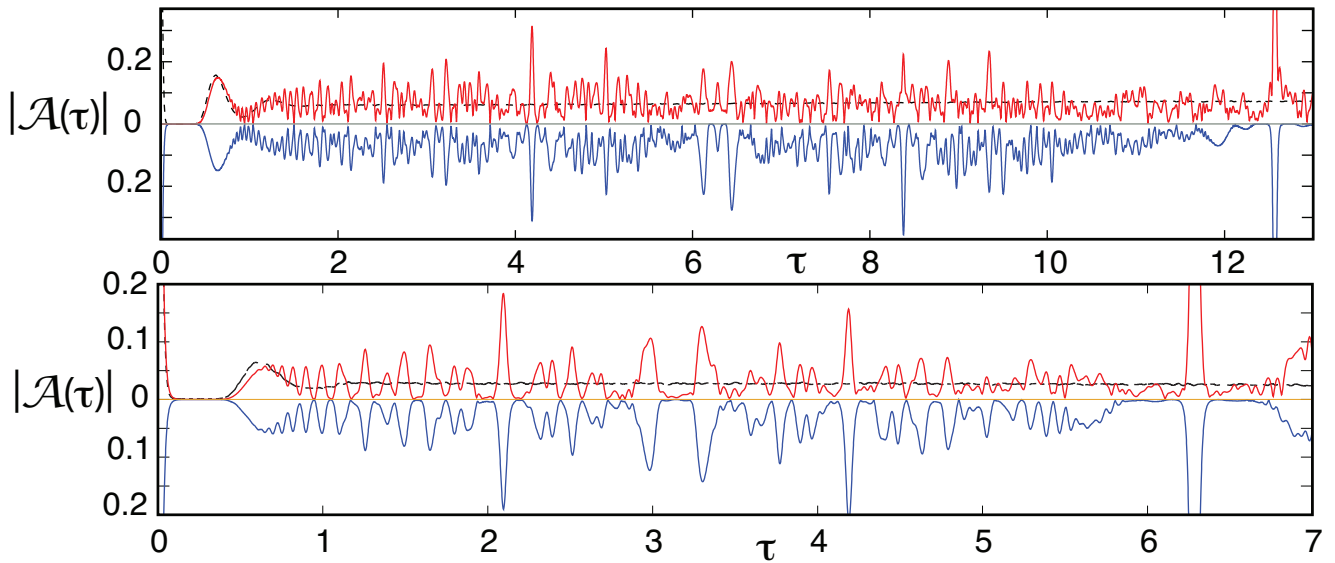}
\caption{(Color online) Comparison of the square root of the quantum time auto-correlation function, Eq.~(\ref{eq:autocor}), with the many-body semiclassical and truncated Wigner approximations.  The solid (blue), dashed (red), and dotted (black) curves represent the quantum, semiclassical, and TWA results, respectively.  The quantum curves are reflected to distinguish the curves better.  The upper panel gives a $4$-site ring example with initial coherent state density wave $|20, 0, 20, 0\rangle$ ($b_j=\sqrt{20}$ or $0$), and values $J=0.2$ and $U=0.5$.   The lower panel gives a $6$-site ring example with initial coherent state density wave $|10, 0, 10, 0, 10, 0\rangle$, and $J=0.2$ and $U=1.0$. The initial decays and revivals (at $t=4\pi, 2\pi$, respectively) are cutoff to expand the scale of $|{\cal A}(\tau)|$. \label{fig1} }
\end{figure*}

The \emph{truncated Wigner approximation} (TWA)~\cite{Steel98,Sinatra02} appears to produce good numerical simulations of Bose gas dynamics through the scrambling time, as long as one focuses on few-body observables. In practice, the TWA samples the initial quantum state of a bosonic many-body system in terms of classical fields, which are numerically propagated via a time-dependent Gross-Pitaevskii equation and summed incoherently. Thus, the TWA lacks an important ingredient, namely the many-body interference effects, which become particularly relevant post-Ehrenfest.  Going beyond TWA without resorting to rather involved numerical ``quantum'' methods based, e.g., on the time-dependent density matrix renormalization group (t-DMRG)~\cite{Daley04, Schollwock05, Schollwock11} or on matrix product states (MPS)~\cite{Verstraete04} (which would fail to reach the mesoscopic regime) requires the implementation of a truly \emph{semiclassical} technique.  It would account for the phases that are associated with the mean-field (MF) trajectories of the classical TWA sampling~\cite{Engl14,Engl15,Simon14,Ullmo18}.  Exploiting the formal similarity between the $N \to \infty$ limit of the bosonic many-body systems and the $\hbar \to 0$ limit of a one-body problem, a proper theory can be constructed for the many-body case by generalizing the time-dependent semiclassical techniques developed in the one-body context~\cite{Maslov81}.

The goal here is to develop this method, illustrate it with a Bose-Hubbard model, and to
show that such an approach, both, quantitatively accounts for quantum many-body interference effects
and qualitatively provides insight into the underlying interference mechanisms.  Furthermore, the
time-energy Fourier transform of the semiclassical dynamics provides detailed spectroscopic
information of the many-body system.  Ahead comparing the semiclassical predictions with those
derived from the classical TWA clearly indicates the onset and presence of many-body quantum interferences post-Ehrenfest. The approach taken is based on the coherent propagation of a Lagrangian manifold (to be defined below) of MF-trajectories.  It is used to identify {\em saddle} MF-trajectories whose classical actions determine the appropriate phases, thereby lifting the time-dependent semiclassical approximation \`a la Maslov~\cite{Maslov81} to many-particle systems. 

Consider Bose-Hubbard systems with tunable tunneling and interaction terms, respectively: 
\begin{equation}
\hat H = -J \sum_{j=1}^N \left(a^\dagger_j a_{j+1} + h.c.\right) +
\frac{U}{2} \sum_{j=1}^N \hat n_j \left(\hat n_j - 1 \right) 
\end{equation}
where $N$ is the number of sites arranged on a ring.  $U$ denotes the strength of the two-body interaction, which depends on the s-wave scattering length of the atomic species considered.  $J$ controls the tunneling amplitude, which depends on the well depth.  Instead of evaluating the evolution of single-particle observables as typically done in methods such as t-DMRG~\cite{Daley04, Schollwock05, Schollwock11} or MPS~\cite{Verstraete04}, our focus is on a more involved observable, i.e.~the evolution of initial states corresponding to {\em coherent states}.  Indeed, because they begin maximally localized with minimum uncertainty, they correspond to the most {\em classical} states, and therefore provide an excellent way to investigate the onset of genuinely quantum effects.  Moreover, they have already been shown to be experimentally relevant in cold-atom physics~\cite{Greiner02b}.  

A challenging initial state to consider for the theory is a {\em coherent state density wave} denoted 
\begin{equation}
|{\bf n}\rangle = \prod_{j=1}^N {\rm e}^
{\left(-\frac{\left|b_j\right|^2}{2} + b_j  \hat a^\dagger_j \right)}|{\bf  0}\rangle = e^{-\mathcal{N}/2} e^{\sqrt{{\cal N}} \hat{\alpha}^\dagger}|{\bf  0}\rangle
\end{equation}
with $\hat{\alpha}^\dagger = \sum_{j=1}^N (b_j/\sqrt{\mathcal{N}}) \hat a^\dagger_j$ and $\mathcal{N} = \sum_{j=1}^N |b_j|^2$, where each site $j$ is loaded with a coherent state of mean particle number $n_j=\left|b_j\right|^2$.  We choose density waves of the form $|n,0,n,0,...,n,0\rangle$.  It describes a perfect Bose-Einstein condensate (in a gauge-symmetry breaking coherent-state representation~\cite{Lieb07}) that populates every other site of the lattice with altogether $\mathcal{N} = n N / 2$ particles~\cite{rem_exp}.  Its time autocorrelation function gives a convenient measure that very strongly exhibits the post-Ehrenfest many-body quantum interferences. It is denoted
\begin{equation}
{\cal C}(t) =\left|{\cal A}(t)\right|^2 \quad , \quad {\cal A}(t) = \langle {\bf n }\left| \hat U(t) \right| {\bf n}\rangle
\label{eq:autocor}
\end{equation}
where $\hat U(t)$ is the unitary time translation operator.  Such phase-sensitive time autocorrelation functions may play an important role in splitting processes in the spirit of Ref.~\cite{Hofferberth07}, where the subsequent recombination of the split atomic clouds depend on their relative phase.  Their explicit experimental detection is within reach using sophisticated single-site atom counting techniques~\cite{Kaufman16}.  This is somewhat analogous to the situation with pump-probe experiments that measured electronic wave packet revivals and fractional revivals in Rydberg atoms~\cite{Yeazell90,Yeazell91}.

The direct comparison of quantum auto-correlation functions along with their time-dependent semiclassical and TWA approximations for 4-site and $6$-site coherent state density waves is illustrated in Fig.~\ref{fig1}.  Note first that there are two relevant time scales,
\begin{equation}
\tau_1 = \frac{2\pi}{U n_j} = 0.63, \  \tau_2 = \frac{2\pi}{U} = 4\pi,\ 2\pi \ (4-site,6-site),
\end{equation}
that come from the Bose-Hubbard model on-site two-body interaction terms only ($J\!=\!0$): $\tau_1$ is a classical scale associated with first return of MF-trajectories; $\tau_2$ is a quantum scale associated with the revival of the initial quantum state~\cite{Greiner02b}.  The TWA is essentially an accurate approximation to roughly $\tau_1$, up to which there exists either zero or one saddle MF-trajectory.  Shortly thereafter, multiple saddle MF-trajectories signify the start of many-body quantum interferences, and the TWA becomes a rather poor approximation for the auto-correlation function.  On the other hand, the semiclassical approximation remains extremely precise to times significantly larger than $\tau_1$.  The distinct peaks at $\tau_2/3$ are the remains of the $1/3$ fractional revivals of the $J=0$ cases, and whereas they are completely missed by TWA, they are perfectly well reproduced by the semiclassical approximations.  In the $4$-site case, it arises as the result of summing the contributions of roughly 60 saddle contributions at any fixed time in its neighborhood. The remains of the full revival at $4\pi=12.57$ is a similar situation, but requires the summation of roughly 600 saddle trajectories; the $6$-site case requires an order of magnitude more.
\begin{figure}[tbh]
\includegraphics[width=8.5 cm]{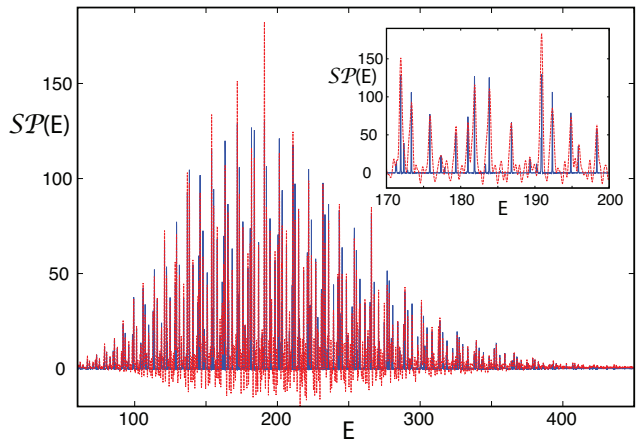}
\caption{(Color online) Comparison of the quantum and semiclassical many-body spectra for the $4$-site example.  To reduce spurious ringing, a Gaussian cutoff was applied to the Fourier transforms with $\sigma=40,6$ for the quantum and semiclassical ${\cal A}(\tau)$, respectively.  Semiclassical theory, good to $O(\hbar^2)$, cannot reproduce the energy centroid exactly.  With a constant energy shift of $E_0=0.9$, the quantum and semiclassical spectra align perfectly, as seen on an enlarged scale in the inset. \label{fig2} }
\end{figure}

The Fourier transform of ${\cal A}(\tau)$ (with phase information) generates detailed spectral information,
\begin{equation}
{\cal SP}(E) = \sum_\nu \left| \langle E_\nu | \vec n \rangle
\right|^2 \delta (E\!-\!E_\nu) \propto \int {\rm d} \tau \ {\rm e}^{iE\tau} {\cal A}(\tau) \, .
\end{equation}
The quantum and semiclassical spectra are compared in Fig.~\ref{fig2}; no spectrum derives from the TWA.  The agreement with the semiclassical theory is excellent.

Derivation of our semiclassical method begins with the quadrature operators $(\hat q_j, \hat p_j)$ defined as 
\begin{equation}
\hat a_j = \left(\hat q_j + i \hat p_j\right)/\sqrt{2} \quad , \quad \hat a_j^\dagger = \left(\hat q_j - i \hat p_j\right)/\sqrt{2} \; .
\end{equation}
For large total particle numbers, the mean field solutions of the Bose-Hubbard model in a phase space representation follow by a substitution of the quadrature operators $(\hat q_j, \hat p_j)$ by
$c$-numbers, which after attention to operator ordering issues results in a classical Hamiltonian
\begin{eqnarray}
\label{hamiltonian}
H_{cl} &=& -J \sum_{j=1}^N \left(q_j q_{j+1} + p_j p_{j+1}\right) + \frac{U}{2}
\sum_{j=1}^N \left(\frac{q_j^2 + p_j^2}{2}\right)^2 \nonumber \\ 
&& - U \sum_{j=1}^N \frac{q_j^2 + p_j^2}{2} \quad .
\end{eqnarray}
The solutions of the resulting Hamilton's equations for various initial conditions up to propagation time, $t$, give the mean field solutions. 

In a $q$-representation, coherent states appear as Gaussian wave packets~\cite{Glauber63}, and in particular 
%\begin{align}
%\langle \vec q| {\bf n}  \rangle &= \pi^{-N/4} \exp\left[- S_{\rm
%    wp}(\vec q) \right] \label{eq:WP} \quad , \nonumber \\
%S_{\rm wp}(\vec q) & = \frac{1}{2} \sum_{j=1}^N  \left(q_j-\sqrt{2n_j}\right)^2  \quad .
%\end{align}
\begin{equation}
 \label{eq:WP} 
\langle \vec q| {\bf n}  \rangle = \pi^{-\frac{N}{4}} \exp\left[- S_{\rm wp}(\vec q) \right], 
\ S_{\rm wp}(\vec q) = \frac{1}{2} \sum_{j=1}^N  \left(q_j-\sqrt{2n_j}\right)^2 .
\end{equation}
This gives rise to the corresponding density operator Wigner transforms,
\begin{equation*}
{\cal W}(\vec q,\vec p) = \left(\frac{1}{\pi}\right)^N \prod_{j=1}^N
\exp \left[ -\left(q_j-\sqrt{2n_j}\right)^2 - p_j^2 \right]  \quad .
\end{equation*}
This density is used for the TWA calculations, which formally are given by solving the evolution equation
\begin{equation} 
\label{eq:TWA}
\frac{\rm d}{{\rm d}t} {\cal W}(\vec q,\vec p) = \left\{H_{cl}, {\cal
    W}(\vec q,\vec p)\right\} \; ,
\end{equation}
with $\{\cdot,\cdot\}$ the Poisson bracket.  Typically, a Monte Carlo weighted sampling of initial conditions is propagated with Hamilton's equations.

Implementation of the semiclassical theory~\cite{Maslov81} can be summarized as follows.  First, the collection of initial conditions (in classical phase space) represented in the initial state's Lagrangian manifold is propagated a time $t$; and all those MF-trajectories whose ending points lie on the final state's Lagrangian manifold give rise to a contribution to the quantity of interest (here
${\cal C}(t)$).  The MF-trajectories satisfying this prescription are saddle MF-trajectories, {\em i.e.}, they are 
associated with a saddle point condition in the overlap integral. For a given saddle MF-trajectory, its
contribution is then expressed in terms of the following quantities: the time integral of the Lagrangian along the trajectory, the Maslov phase index~\cite{Maslov81}, and the
stability matrix $\bf M$ describing the linearized motion near the trajectory.   

Generically, the Lagrangian manifold associated with a semiclassical  wavefunction $\Psi_{\rm sc}(\vec{q}) = a(\vec{q}) \exp [i S_{\rm sc}(\vec{q})]$ is given by $\vec{\cP} (\vec{\cQ}) =  \vec{\nabla} S_{\rm sc}(\vec{\cQ})$.  A coherent state, Eq.~\eqref{eq:WP}, is indeed in the usual semiclassical form with the peculiarity that $S_{\rm sc} = i S_{\rm WP}$ is an imaginary function. The Lagrangian manifolds of the initial and final states, respectively, are thus given by the relations~\cite{Huber88}
\begin{equation*}
{\cQ}_j - \sqrt{2n_j} = - i  {\cP}_j  \ \mbox{(initial)} \, , \, {\cQ}_j -\sqrt{2n_j} = +  i  {\cP}_j   \ \mbox{(final)} 
\end{equation*}
for each value of $j$ where the phase space coordinates (${\cQ}_j,{\cP}_j$) and Hamiltonian, 
Eq.~(\ref{hamiltonian}), have been analytically continued to complex variables.  Note that this approach is quite general~\cite{Maslov81}.  Fock states could be treated in roughly the analogous way as coherent states.  For occupied sites, the Lagrangian manifold of a Fock state would involve only real trajectories.  However, since for unoccupied sites, the Fock state and coherent state are identical, a hybrid method with a Lagrangian manifold relying on real initial conditions for occupied sites and complex initial conditions for initially empty sites would result. 

For semiclassical coherent state propagation, the need to work with complexified phase space variables comes with a host of its own technical challenges~\cite{Huber88, Baranger01}.  At the most basic level, one must find the saddle trajectories, which in a high dimensional complex phase space is nontrivial, and second one must know which saddle trajectories must be thrown away because of the square integrability boundary condition.  An approach to the latter problem is given in~\cite{Pal16} following the work of~\cite{Tomsovic91b, Oconnor92, Tomsovic93, Barnes94}.  In short, the complex set of saddles that contribute can be put in a one-to-one correspondence with contributions of real classical transport pathways, and a Newton-Raphson algorithm locates the complex saddle trajectory for each pathway.  That leaves finding a practical solution to locating these transport pathways.  It requires understanding the asymptotic structure and flow in phase space.  Certain directions lead to maximal exploration, whereas others lead to none.  Identifying the relevant directions allows one to reduce greatly the dimensionality of the search space,  in fact to just a few dimensions~\cite{Kocia14b,Kocia15,Tomsovic18}.
 
Once the relevant saddle MF-trajectories $\gamma$ have been identified, the
computation of the various quantities needed (action integral $S(\vec
{\cQ}_t^\gamma,\vec {\cQ}_0^\gamma;t)$, Maslov index
$\nu_\gamma$, and stability matrix ${\bf M}^\gamma$) is essentially
straightforward.  Introducing the ``scaled'' time variable $\tau
\equiv t/\hbar$,  and the corresponding scaled action 
$S(\vec{\cQ}_\tau^\gamma, \vec{\cQ}_0^\gamma;\tau) = \hbar^{-1}
S(\vec{\cQ}_\tau^\gamma, \vec{\cQ}_0^\gamma;t)$, 
the auto-correlation function ${\cal C}(\tau)$ can be expressed as~\cite{Huber88}
\begin{align}
\label{eq:autocor-sc}
 {\cal C}(\tau) &= \left|\sum_\gamma c^{1/2}_\gamma(\tau) \exp \left[ i
      \phi_{\gamma}(\tau) \right]  \right|^2 \; ,  \\
i \phi_{\gamma}(\tau) &= i S(\vec{\cQ}_\tau^\gamma, \vec{\cQ}_0^\gamma;\tau) - i \nu_\gamma
        \frac{\pi}{2} + F^{\gamma-}_0 + F^{\gamma,+}_\tau \; , \nonumber \\
 c_\gamma(\tau) & = 
        {\rm Det}^{-1}\left[\frac{1}{2} 
          ({\bf M_{11}^\gamma} + {\bf M_{22}^\gamma} + i {\bf M_{21}^\gamma}  -
          i{\bf M_{12}^\gamma}) \right] 
  \; , \nonumber
\end{align}
with 
\begin{align}
F^{\gamma,-}_0 &= i \vec {\cP}_0^R \cdot \vec{\cP}_0^I 
- \frac{1}{2} \vec {\cP}_0^I \cdot \vec {\cP}_0^I - \frac{1}{2} \vec {\cQ}_0^I \cdot \vec {\cQ}_0^I -
 \vec {\cP}^R_0 \cdot \vec {\cQ}^I_0 \, , \nonumber \\
F^{\gamma,+}_\tau &= i \vec{\cP}_\tau^R \cdot \vec{\cP}_\tau^I - 
\frac{1}{2} \vec{\cP}_\tau^I \cdot \vec{\cP}_\tau^I -
\frac{1}{2} \vec {\cQ}_\tau^I \cdot \vec {\cQ}_\tau^I + 
 \vec{\cP}^R_\tau \cdot \vec {\cQ}^I_\tau
\end{align} 
(where $R,I$ refer to real and imaginary parts) and 
\begin{equation}
{\bf M_{11}^\gamma} = \frac{\partial \vec{\cQ}_\tau}{\partial
  \vec{\cQ}_0} \, , \,  {\bf M_{22}^\gamma} = \frac{\partial \vec{\cP}_\tau}{\partial
  \vec{\cP}_0} \, , \,  {\bf M_{12}^\gamma} = \frac{\partial \vec{\cQ}_\tau}{\partial
  \vec{\cP}_0} \, , \,  {\bf M_{21}^\gamma} = \frac{\partial \vec{\cP}_\tau}{\partial
  \vec{\cQ}_0}.
\nonumber
  \end{equation}

Interestingly, the saddle MF-trajectories provide an alternative way of producing the TWA.  In the limit of large particle number and an ``infinitely'' dense Monte Carlo, the approximation~\eqref{eq:TWA} is asymptotically equivalent to the diagonal contribution in Eq.~(\ref{eq:autocor-sc}), $ {\cal C}_{\rm diag}(\tau)  \equiv \sum_\gamma c_\gamma(\tau)$.

In summary, the semiclassical approach to mesoscopic many-body quantum dynamics produces an accurate approximation with fully incorporated many-body interferences.  It is developed here as a practical technique, i.e.~the technical problems of implementation are solvable.  It is best adapted to systems with mesoscopic populations of particles or more, and most accurate over short to intermediate time scales.  The semiclassical approximation being effectively an expansion in the inverse of the density, the accuracy improves with increasing particle number.  Furthermore, its accuracy is capable of giving rather detailed many-body spectroscopic information.  In stark contrast, the TWA smooths over all the post-Ehrenfest many-body quantum interferences and cannot provide spectroscopic information. 

The relaxation dynamics of a mesoscopically populated, coherent state density wave in the strong interaction regime provided both a physically interesting and stringent test of the theory.  The remnants of matter revivals occur in its quantum dynamics, and the reconstruction of their presence is highly non-trivial.  In addition, this density wave gives rise to a spectrum that is quite reminiscent of those found in various cases of high-resolution molecular spectroscopy~\cite{Heller81b, Gruebele92}.

The semiclassical method is extremely general, and leads naturally to many possible lines of future research.  For example, it can be adapted to the dynamics of a variety of different initial states.  It can also be adapted to other measures besides autocorrelation functions, such as fidelities and out-of-time-ordered correlators, which are so important in quantum information studies.  The method also applies equally well to the study of all interaction regimes; i.e.~the saddle trajectories can be found for any dynamical regime, chaotic or not.  However, for chaotic dynamical systems, the exponential proliferation of saddles may shorten the time scale of practical application.  Our semiclassical calculations have also been performed for $8$-site rings with up to 160 particles, presumably beyond the possibility of full quantum calculations.  With a bit more effort, they could be extended to greater numbers of sites, and other configurations besides the rings considered here.  Finally, we mention that with the help of a semiclassical propagator in fermionic Fock space (such as proposed in Ref.~\cite{Engl14}), the present method can be extended to the case of fermionic atoms. This opens various perspectives for studying the interplay of scrambling and localization phenomena in the context of ultracold (bosonic or fermionic) gases.

S.T.\ acknowledges support from the Vielberth Foundation and the UR International Presidential Visiting Fellowship 2016 during two extended stays at the Physics Department of Regensburg University.

\bibliography{quantumchaos,general_ref,manybody,molecular}

\begin{thebibliography}{54}
\bibitem [{rem()}]{rem_exp}%
  \BibitemOpen
  \href@noop {} {}\bibinfo {note} {Experimentally, such condensate states can,
  for $N=4$, possibly be prepared by loading ultracold bosonic atoms into
  (transversally slightly elongated) plaquettes of a two-dimensional optical
  lattice. Using standard superlattice techniques~\cite{Folling07}, the
  energies on two opposite plaquette sites can be manipulated during this
  preparatory stage such that the atoms settle to a $|n,0,n,0\rangle$
  configuration.}\BibitemShut {Stop}%
\end{thebibliography}

\end{document}